# Ascorbic acid enhances the inhibitory effect of aspirin on neuronal cyclooxygenase-2-mediated prostaglandin $E_2$ production

Eduardo Candelario-Jalil [a], Ravi S. Akundi [a], Harsharan S. Bhatia [a], Klaus Lieb [a],

Kurt Appel [b], Eduardo Muñoz [c], Michael Hüll [a], Bernd L. Fiebich [a, b, *]

[a]Neurochemistry Research Group, Department of Psychiatry, University of Freiburg Medical School, Hauptstrasse 5, D-79104 Freiburg, Germany

[b]VivaCell Biotechnology GmbH, Ferdinand-Porsche-Str. 5, D-79211 Denzlingen, Germany

[c]Departamento de Biología Celular, Fisiología e Inmunología. Universidad de Córdoba, Avda Menéndez Pidal s/n. 14004, Córdoba, Spain

* Corresponding author:
Bernd L. Fiebich, Ph.D.
Department of Psychiatry
University of Freiburg Medical School
Hauptstrasse 5
D-79104 Freiburg, Germany
Tel. (49)-761-270-6898
Fax (49)-761-270-6917
E-mail: bernd.fiebich@klinikum.uni-freiburg.de

**Running title**: Inhibition of neuronal prostaglandin synthesis by ascorbic acid

**Abstract**

Inhibition of neuronal cyclooxygenase-2 (COX-2) and hence prostaglandin $E_2$ ($PGE_2$) synthesis by non-steroidal anti-inflammatory drugs has been suggested to protect neuronal cells in a variety of pathophysiological situations including Alzheimer's disease and ischemic stroke. Ascorbic acid (vitamin C) has also been shown to protect cerebral tissue in a variety of experimental conditions, which has been attributed to its antioxidant capacity. In the present study, we show that ascorbic acid dose-dependently inhibited interleukin-1β (IL-1β)-mediated $PGE_2$ synthesis in the human neuronal cell line, SK-N-SH. Furthermore, in combination with aspirin, ascorbic acid augmented the inhibitory effect of aspirin on $PGE_2$ synthesis. However, ascorbic acid had no synergistic effect along with other COX inhibitors (SC-58125 and indomethacin). The inhibition of IL-1β-mediated $PGE_2$ synthesis by ascorbic acid was not due to the inhibition of the expression of COX-2 or microsomal prostaglandin E synthase (mPGES-1). Rather, ascorbic acid dose-dependently (0.1-100 μM) produced a significant reduction in IL-1β-mediated production of 8-*iso*-prostaglandin $F_{2\alpha}$ (8-*iso*-$PGF_{2\alpha}$), a reliable indicator of free radical formation, suggesting that the effects of ascorbic acid on COX-2-mediated $PGE_2$ biosynthesis may be the result of the maintenance of the neuronal redox status since COX activity is known to be enhanced by oxidative stress. Our results provide *in vitro* evidence that the neuroprotective effects of ascorbic acid may depend, at least in part, on its ability to reduce neuronal COX-2 activity and $PGE_2$ synthesis, owing to its antioxidant properties. Further, these experiments suggest that a combination of aspirin with ascorbic acid constitutes a novel approach to render COX-2 more sensitive to inhibition by aspirin, allowing an anti-inflammatory therapy with lower doses of aspirin, thereby avoiding the side effects of the usually high dose aspirin treatment.

**Introduction**

A large body of evidence implicates both brain inflammation and oxidative stress in the pathophysiological mechanisms contributing to the neuronal death seen in different neurological disorders including Alzheimer's disease (AD), Parkinson's disease and stroke (for a recent review, see Andersen, 2004). Inflammatory processes may contribute to the neurodegenerative process through mechanisms involving the release of pro-inflammatory cytokines, reactive oxygen species (ROS), nitric oxide (NO) and prostaglandins.

Prostaglandins are produced by the enzymatic action of cyclooxygenase (COX). Two isoforms of COX have been described: COX-1 and COX-2 (Seibert et al., 1997). COX-1 is expressed constitutively in many organs and contributes to the synthesis of prostanoids involved in normal cellular functions (Seibert et al., 1997). COX-2 is expressed in several cell types after inflammatory stimulation including monocytes and macrophages (Pairet and Engelhardt, 1996). In the CNS, COX-2 is expressed under normal conditions in neurons of the cortex, hippocampus and amygdala (Yamagata et al., 1993; Chang et al., 1996), and is mainly linked to synaptic activity (Kaufmann et al., 1996). Seizure activity and spreading depression induce COX-2 expression while tetrodotoxin or blockade of N-methyl-D-aspartate (NMDA) receptors with MK-801 reduce neuronal COX-2 expression (Yamagata et al., 1993; Miettinen et al., 1997).

COX-2 has become the focus of attention because it is the rate-limiting enzyme involved in arachidonic acid metabolism, thereby generating prostaglandins and thromboxanes, molecules that play important roles in supporting and sustaining the inflammatory response (Vane et al., 1998), and the induction of COX-2 results in an inflammatory cascade accompanied by formation of ROS.

In both, AD and cerebral ischemia, neuronal COX-2 expression may exaggerate CNS damage. In AD, elevated expression of COX-2 mRNA and protein has been reported (Oka and Takashima, 1997;

Pasinetti and Aisen, 1998). COX-2 protein expression may be elevated in hippocampal neurons in AD although findings have been contradictory (Chang et al., 1996; Ho et al., 1999). In animal models of hypoxia and in human ischemia, COX-2 is abundantly expressed in neurons, astroglia and microglia (Ohtsuki et al., 1996). Elevated neuronal COX-2 expression in CNS pathology may not depend on synaptic activity but on other inducers of COX-2 such as the proinflammatory cytokine interleukin-1β (IL-1β) (Harris et al., 1996; Fiebich et al., 2000b; Touzani et al., 2002). In transgenic animals overexpressing COX-2, glutamatergic excitotoxicity is enhanced (Kelley et al., 1999). COX-2 deficient mice *vice versa* show a reduced neuronal vulnerability to ischemia and excitotoxicity (Iadecola et al., 2001).

Several epidemiological studies suggested a strong protective effect of nonsteroidal anti-inflammatory drugs (NSAIDs) against the development of AD (in t'Velt et al., 2001). The anti-inflammatory, analgesic and antipyretic actions of NSAIDs such as aspirin are thought to be mainly due to an inhibition of COX-2 leading to a lowered prostaglandin synthesis (Pairet and Engelhardt, 1996).

Ascorbic acid (vitamin C) is an antioxidant vitamin which accumulates in the brain from the blood supply and is maintained at high concentrations in both neuronal and glial cells (Rebec and Pierce, 1994; Rice and Russo-Menna, 1998). Ascorbic acid serves as a neuromodulator in normal brain functioning and as a neuroprotective agent in various conditions of oxidative stress (Grunewald, 1993). The neuroprotective action of ascorbic acid has been mainly attributed to its antioxidant capacity.

In the present study, we tested a broader field of possible functions of ascorbic acid and investigated whether ascorbic acid is able to inhibit COX-2-mediated $PGE_2$ synthesis and may augment the inhibitory effect of aspirin on PGE$_2$ synthesis. As a model system, we used the human neuronal cell line SK-N-SH. In these cells, COX-2 and subsequent prostaglandin synthesis can be induced by IL-1β, a cytokine which is involved in neuroinflammation (Harris et al., 1996; Rothwell and Luheshi, 2000; Fiebich et al., 2000b; Touzani et al., 2002). The present study was prompted by our previous findings, which showed a potent inhibitory effect of ascorbic acid on PGE$_2$ production in lipopolysaccharide (LPS)-activated primary rat microglial cells (Fiebich et al., 2003).

**Materials and Methods**

**Materials**

Human IL-1β, which is effective on mouse and human cells, was purchased from Roche Diagnostics (Mannheim, Germany). L-Ascorbic acid, Dehydro-L-(+)-Ascorbic acid (DHA), Trolox C (6-hydroxy-2,5,7,8-tetramethylchroman-2-carboxylic acid), indomethacin, and aspirin (acetylsalicylic acid) were obtained from Sigma-Aldrich (Deissenhofen, Germany). The highly selective COX-2 inhibitor SC-58125 (1-[(4-methysufonyl)phenyl]-3-tri-fluoromethyl-5-(4-fluorophenyl) pyrazole) and the highly selective COX-1 inhibitor SC-560 (5-(4-chlorophenyl)-1-(4-methoxyphenyl)-3-(trifluoromethyl)-1H-pyrazole) were obtained from Cayman Chemical (Ann Arbor, Michigan, USA). All compounds, used at the given concentrations, do not affect the viability of the cells as observed through the CellTiter-Glo® luminescent cell viability assay kit (Promega, Mannheim, Germany), which measures metabolic ATP levels (data not shown).

**Cell culture**

The human neuroblastoma cell line SK-N-SH was obtained from the American Type Culture Collection (HTB-11, Rockville, USA) and was grown in MEM-Earle's medium (PAA, Cölbe, Germany), which does not contain any ascorbic acid. Medium was supplemented with 5% fetal calf serum (FCS) (PAN, Aidenbach, Germany), 2 mM L-glutamine, 1 mM sodium pyruvate, 40 units/ml penicillin/streptomycin (all purchased from PAA Laboratories, Cölbe, Germany), 0.4% MEM vitamins (not containing ascorbic acid) and 0.4% MEM non-essential amino acids (both purchased from Invitrogen GmbH, Karlsruhe, Germany). Confluent monolayers were passaged routinely by trypsinization. Cells were plated for protein and RNA analysis in 6-well plates ($2.8 \times 10^5$ cells/well) and for determination of $PGE_2$ and 8-*iso*-prostaglandin $F_{2\alpha}$ (8-*iso*-$PGF_{2\alpha}$) in 24-well plates ($9 \times 10^4$ cells/well). Cultures were grown at 37°C in 5% $CO_2$ until 80% confluence, and the medium was changed the day before treatment.

**Prostaglandin $E_2$ ($PGE_2$) Enzyme immunoassay**

SK-N-SH cells were plated in 24-well cell culture plates and pre-incubated for 30 min with the inhibitors to be tested. Thereafter, cells were treated with IL-1β (10 U/ml) for 24 h. Supernatants were then used for the determination of $PGE_2$. Supernatants were harvested, centrifuged at 10,000 x g for 10 min and levels of $PGE_2$ in the media were measured by enzyme immunoassay (EIA) (Biotrend, Köln, Germany) according to the manufacturer's instructions. Standards from 39 to 2500 pg/ml were used; sensitivity of the assay is 36.2 pg/ml.

**Cyclooxygenase (COX) activity assay**

To determine any direct inhibitory effect of ascorbic acid, DHA or the combination of aspirin with ascorbic acid on COX enzymatic activity, an arachidonic acid assay was performed (Fiebich et al., 2003). Briefly, SK-N-SH cells were plated in 24-well cell culture plates and pre-incubated with IL-1β (10 U/ml) for 24 h. Medium was then changed to a serum-free medium. Inhibitors were added for 15 min, and arachidonic acid (15 μM final concentration) was supplemented for another 15 min. Supernatants were then collected for the determination of $PGE_2$ as described above.

**Polyacrylamide gel electrophoresis and Western blot analysis**

SK-N-SH cells were left untreated or treated with IL-1β (50 U/ml) in the presence or absence of ascorbic acid (100 µM), aspirin (100 µM) and a combination of both (each at 100 µM) for 24 h. Cells were then washed with Dulbecco's phosphate buffered saline (Cell Concepts GmbH, Umkirch, Germany) and lysed in 1.3 % SDS (sodium dodecyl sulfate) containing 100 µM orthovanadate (Laemmli, 1970). Lysates were homogenized by repeated passage through a 26-gauge needle. Protein contents were measured using the bicinchoninic acid method (BCA protein determination kit from Pierce, distributed by KFC Chemikalien, München, Germany) according to the manufacturer's instructions using bovine serum albumin (BSA, Sigma) as a standard (ranging from 2 to 40 µg/µl; optical density read at 570 nm). Before electrophoresis, 0.1% bromophenol blue and 10 mM DTT were added to the samples and incubated at 95°C for 5 min before loading. For western blotting, 40 µg of protein from each sample was subjected to SDS-PAGE (polyacrylamide gel electrophoresis) on a 7.5% gel under reducing conditions. Proteins were then transferred onto a polyvinylidene fluoride (PVDF) membrane (Millipore, Bedford, MA, USA) by semi-dry blotting. The membrane was blocked overnight at 4°C using Rotiblock (Roth, Karlsruhe, Germany) and for another hour at room temperature before incubation with the primary antibody. Primary antibodies were goat anti-COX-2 (M-19) or goat anti-COX-1 (M-20) obtained from Santa Cruz Biotechnology (Heidelberg, Germany). Both antibodies were diluted 1:500 in Tris-buffered saline (TBS, 50 mM, pH 8.0) containing 0.1% Tween 20 (Merck, Darmstadt, Germany) and 1 % bovine serum albumin (BSA, Sigma). Membranes were incubated with the corresponding primary antibody for 2 h at room temperature. After extensive washing (three times for 15 min each in TBS containing 0.1% Tween 20), proteins were detected with horse radish peroxidase-coupled rabbit anti-goat IgG (Santa Cruz, 1:100,000) using chemiluminescence (ECL) reagents (Amersham Pharmacia Biotech, Freiburg, Germany). Quantification of the Western Blots was performed using ScanPack 3.0 software (Biometra, Göttingen, Germany). Equal protein loading and transfer were assessed by subjection of each sample to a Western blot for actin (rabbit anti-actin antibody diluted 1:1000, obtained from Sigma, Saint Louis, MO, USA). All western blot experiments were carried out at least three times.

**RNA extraction and reverse transcription-polymerase chain reaction (RT-PCR)**

Total RNA was extracted using the guanidine isothiocyanate method according to Chomczynski and Sacchi (Chomczynski and Sacchi, 1987). For RT-PCR, 2 µg of total RNA was reverse transcribed using Moloney Murine Leukemia Virus (M-MLV) reverse transcriptase (Promega, Mannheim, Germany), RNase Inhibitor rRNasin® (Promega), dNTP master mix (Invitek, Berlin, Germany) and random hexamer primers (Promega). PCR was carried out using *Taq* DNA polymerase (Promega), dNTP master mix (Invitek, Berlin, Germany) and the following primers: human microsomal prostaglandin E synthase-1, mPGES-1 (forward: 5'-CTCTGCAGCACGCTGCTGG-3', reverse: 5'-GTAGGTCACGGAGCGGATGG-3', annealing temperature 65°C, 35 cycles, amplicon size: 338 bp). Equal equilibration was determined using human β-actin primers (forward: 5'-ATCTGGCACCACACCTTCTACAATGAGCTGCG-3', reverse: 5'-CGTCATACTCCTGCTTGCTGATCCACATCTGC-3', 60°C, 30 cycles, product length: 838 bp). PCR products were separated electrophoretically on a 2% agarose gel. Potential contamination by genomic DNA was controlled by omitting reverse transcriptase and using β-actin primers in the subsequent PCR amplification. Only RNA samples showing no bands after this procedure were used for further investigation. Primers were designed using the Primer 3 software developed by the Whitehead Institute for Biomedical Research (http://frodo.wi.mit.edu/primer3/primer3_code.html) and synthesized through an in-house facility (Dr. Gabor Igloi, Institute for Biology III, Freiburg, Germany).

**Determination of 8-*iso*-prostaglandin $F_{2\alpha}$ (8-*iso*-$PGF_{2\alpha}$)**

SK-N-SH cells were pre-treated for 30 min with different concentrations of ascorbic acid (0.1-100 µM), DHA (0.1-100 µM) or Trolox C (10 and 100 µM), or with aspirin (0.1-100 µM; alone or in combination with 10 µM ascorbic acid), indomethacin or SC-58125. After 30 min pre-stimulation, cells were stimulated with 10 U/ml IL-1β for 24 h. Supernatants were harvested and the levels of 8-*iso*-$PGF_{2\alpha}$ were measured by an enzyme immunoassay according to the manufacturer's instructions (Cayman Chemicals, Ann Arbor, MI, USA). The standards were used in the range of 3.9 to 500 pg/ml (detection limit of 5 pg/ml).

**Data Analysis**

Data from at least 3 experiments with 2-3 wells per concentration (n=8-9) were used for data analysis. Statistical analysis was done by the Institut für Biochemische Analysen und Methodenentwicklung GbR (Freiburg, Germany). Original data were converted into %-values of IL-1β control and mean ± S.D. calculated. Inhibitor concentration at which 50% inhibition is observed (defined as $IC_{50}$) was calculated by computerized non-linear regression analysis. In the figures, a regression curve was adjusted to the calculated mean values. In some experiments, data are presented as mean ± S.D. and values were compared using *t*-test (two groups) or one-way ANOVA with *post-hoc* Student-Newman-Keuls test (multiple comparison).

**Results**

The effects of ascorbic acid and aspirin on $PGE_2$ production were investigated in the human neuroblastoma cell line SK-N-SH. Basal production of $PGE_2$ in these cells (i.e. during unstimulated conditions) was 138 pg/ml ± 101 pg/ml (mean ± S.D.). The addition of increasing amounts of IL-1β led to a dose-dependent increase in COX-2 protein expression with the concomitant $PGE_2$ biosynthesis, which is maximal at 10-50 U/ml IL-1β as shown in our previous report (Fiebich et al., 2000b). After 24 h (IL-1β, 50 U/ml), $PGE_2$ production in SK-N-SH cells increased approximately 4-fold to 637 pg/ml ± 145 pg/ml (mean ± S.D.).

As expected, aspirin dose-dependently inhibited IL-1β-induced $PGE_2$ synthesis with an $IC_{50}$ of 2.32 μM (Fig. 1A). Ascorbic acid alone also caused a dose-dependent inhibition of IL-1β-induced $PGE_2$ biosynthesis, starting at 1 μM (20% inhibition). The maximal effect was an inhibition of approximately 62.5% at a dose of 100 μM with an $IC_{50}$ of 6.76 μM (Fig. 1A). In further experiments, increasing concentrations of ascorbic acid were added to aspirin to investigate a possible additive effect of both substances. As shown in Fig. 1A, a possible synergistic effect of ascorbic acid (10, 30 and 100 μM) was observed, mainly in combination with low doses of aspirin (10 and 100 nM). The combination of aspirin with 10 μM ascorbic acid resulted in a significant decrease in the $IC_{50}$ for aspirin, going from 2.32 μM (aspirin alone) to 0.85 μM (Fig. 1A). At high doses of ascorbic acid (30 and 100 μM), the inhibitory effect of aspirin was potently enhanced resulting in an $IC_{50}$ of 0.68 and 0.15 μM, respectively (Fig. 1A). Thus, in this system, ascorbic acid produced a synergistic action on aspirin-mediated inhibition of $PGE_2$ production. In other words, the combination of aspirin and ascorbic acid (100 μM) was 15 times more potent than aspirin alone in inhibiting IL-1β-induced $PGE_2$ synthesis in neuronal cells. Interestingly, Trolox C, an antioxidant used as control, potently reduced $PGE_2$ production in SK-N-SH cells stimulated with IL-1β as shown in Fig. 1B. The addition of indomethacin (a non-selective COX inhibitor used as control) completely prevented IL-1β-mediated $PGE_2$ production (Fig. 1B), confirming our previous observations (Fiebich et al., 2000b).

These initial results raised two main questions: 1) is the inhibitory effect of aspirin on $PGE_2$ production also potentiated by other antioxidants, and 2) does ascorbic acid synergize with other COX inhibitors?. In an attempt to respond to the first question, we conducted an experiment in which several doses of aspirin were combined with Trolox C (1 and 10 µM). As shown in Fig. 2, Trolox C (10 µM) also enhanced the inhibitory effect of aspirin on IL-1β-induced $PGE_2$ synthesis in SK-N-SH cells. A clear example of this possible synergistic effect is observed at the lower dose: aspirin at 0.1 µM had no effect on IL-1β-mediated $PGE_2$ production, but in combination with 10 µM Trolox C, a 47% inhibition was observed (p<0.01). Similar effects were also observed at higher doses of aspirin, ranging from 1 to 100 µM (Fig. 2).

The answer to our second question is depicted in Fig. 3. We included other COX inhibitors in this experiment, and interestingly, we didn't observe any additive effect of ascorbic acid on COX-mediated $PGE_2$ formation when combined with these compounds. As expected, the highly selective COX-2 inhibitor SC-58125 significantly reduced $PGE_2$ production, but the combination of ascorbic acid with this compound had no further inhibitory effect on $PGE_2$ synthesis (Fig. 3A). Similar results were found in the case of the non-selective COX inhibitor indomethacin (Fig. 3B).

Since COX-2 expression is in part dependent on redox-sensitive transcriptional events (Fiebich et al., 2000b; Kim et al., 2000; Madrigal et al., 2003), and previous evidences indicate that ascorbic acid reduces COX-2 expression under some conditions (Sanchez-Moreno et al., 2003; Han et al., 2004), we decided to investigate if ascorbic acid is able to affect COX-2 expression in our *in vitro* model. Densitometric analysis of 3 independent Western blots revealed that ascorbic acid and aspirin alone or in combination (100 μM each) did not influence IL-1β-induced synthesis of COX-2 protein in SK-N-SH cells (Fig. 4A and 4B). The non-inducible COX-1 protein synthesis was not affected by IL-1β, ascorbic acid or aspirin (Fig. 4A and 4B).

In our next experiments, we explored an alternative or additional possible effect of ascorbic acid, its oxidized form DHA, and the combination of aspirin and ascorbic acid on the expression of the inducible prostaglandin E synthase (mPGES-1), an important enzyme downstream of COX in the biosynthesis of $PGE_2$. Stimulation of various cultured cells with pro-inflammatory stimuli leads to a marked elevation of mPGES-1, often with concomitant induction of COX-2 (Jakobsson et al., 1999; Murakami et al., 2000; Mancini et al., 2001; Murakami et al., 2002). In this study, we show for the first time a marked upregulation in the expression of mPGES-1 in SK-N-SH cells exposed to IL-1β (10 U/ml) for 4 h (Fig. 5). This pronounced effect of IL-1β was not prevented by any concentration of ascorbic acid, DHA, the combination of aspirin with ascorbic acid or either compound alone, as depicted in Fig. 5.

Considering that a mechanism involving reduction of COX-1 or COX-2 protein expression cannot explain the inhibitory effects of ascorbic acid on IL-1β-induced $PGE_2$ synthesis, and its possible synergistic effect in combination with aspirin (Fig. 4), we decided to investigate whether ascorbic acid might be acting through direct inhibition of COX activity, and/or directly potentiating the inhibitory action of aspirin. Table 1 shows the findings from the COX enzymatic activity assay. At doses of 10 and 100 μM, aspirin produced a 23 and 48 % inhibition of COX activity, respectively. The combination of aspirin with ascorbic acid (10 and 100 μM) did not further decrease aspirin-mediated COX inhibition, as reflected in the levels of $PGE_2$. Similarly, ascorbic acid was unable to directly inhibit COX activity in this assay (Table 1). We also checked if the oxidized metabolite of ascorbic acid, DHA, could modify COX catalysis. Similar to what was found for ascorbic acid, DHA failed to inhibit COX activity (data not shown).

Results from the COX activity assay showed that ascorbic acid is unable to directly inhibit COX activity (Table 1). Control experiments with indomethacin and SC-58125 (a COX-2 selective inhibitor) showed a potent inhibitory effect of both compounds in this COX enzymatic assay (Table 1), confirming the validity of this system to study the inhibitory properties of ascorbic acid on neuronal COX activity. A control experiment with the highly selective COX-1 inhibitor SC-560 (10 and 100 nM) showed that COX-1 inhibition fails to significantly reduce $PGE_2$ in this assay, as presented in Table 1.

After having excluded the possibility that the effects of ascorbic acid in the present study were mediated by downregulation of the expression of key enzymes involved in $PGE_2$ production (COX-1/2, mPGES-1), and taking into account that COX enzymatic activity increases under conditions of oxidative stress due to the activation of COX by hydroperoxides (Hemler and Lands, 1980; Kulmacz and Lands, 1983), we then studied the effects of ascorbic acid and its oxidized metabolite DHA on the formation of 8-*iso*-$PGF_{2\alpha}$, an index of oxidant stress-mediated production of lipid hydroperoxides / ROS. 8-isoprostanes are formed in response to free radical attack on arachidonic acid on membrane phospholipids and are considered as a reliable and highly sensitive measure of reactive oxygen species (ROS) formation (Pratico et al., 1998). We also investigated if the simultaneous incubation with ascorbic acid and aspirin produces an additive effect on IL-1β-induced 8-*iso*-$PGF_{2\alpha}$ production.

IL-1β (10 U/ml) produced a dramatic increase in 8-*iso*-$PGF_{2\alpha}$ concentrations in the culture media collected from SK-N-SH cells treated for 24 h with this pro-inflammatory cytokine (Fig. 6), which is in line with findings of a previous study in neuronal cells (Igwe et al., 2001). Interestingly, ascorbic acid dose-dependently reduced IL-1β-induced oxidative stress as shown in Fig. 6A. Only the highest dose of DHA (100 µM) was able to reduce 8-*iso*-PGF$_{2\alpha}$ formation by ~50%. Treatment with the antioxidant Trolox C potently reduced IL-1β-mediated ROS formation, as evidenced by the maintenance of 8-*iso*-PGF$_{2\alpha}$ concentrations to basal levels with the highest dose (100 µM) of Trolox C (Fig. 6A).

Aspirin dose-dependently inhibited the formation of 8-*iso*-PGF$_{2\alpha}$, an effect that was potentiated by ascorbic acid at the highest doses of aspirin tested (10 and 100 µM), as shown in Fig. 6B. The effect of the combination of low doses (0.1 and 1 µM) of aspirin and ascorbic acid on 8-*iso*-PGF$_{2\alpha}$, was mainly due to the effects of ascorbic acid since no statistically significant differences ($p>0.05$) were observed among low doses (0.1 and 1 µM) of aspirin plus ascorbic acid and ascorbic acid alone (Fig. 6B). Nevertheless, when higher doses of aspirin were tested, the combination of this COX inhibitor with ascorbic acid produced a significant reduction in 8-*iso*-PGF$_{2\alpha}$, which was statistically different ($p<0.05$) from ascorbic acid alone (Fig. 6B).

Since COX activity is associated to the formation of free radicals (Kukreja et al., 1986; Jiang et al., 2004) and there is a dramatic increase in COX-2 enzymatic activity in neuronal cells upon stimulation with the pro-inflammatory cytokine IL-1β, we checked the possibility that COX inhibition might reduce neuronal oxidative stress. Interestingly, our experiments demonstrated potent inhibitory effects of the highly COX-2 inhibitor SC-58125 and the non-selective COX inhibitor indomethacin on 8-*iso*-PGF$_{2\alpha}$ production in IL-1β-stimulated neuronal cells (Fig. 6C).

**Discussion**

This study investigated, for the first time, the inhibitory effect of ascorbic acid alone or in combination with aspirin on $PGE_2$ synthesis in neuronal cells, using the human neuronal cell line SK-N-SH. Our results demonstrate that ascorbic acid mediates moderate inhibition of neuronal $PGE_2$ biosynthesis and that in combination with aspirin leads to a super-additive and almost complete inhibition of COX-2-mediated $PGE_2$ formation. These findings are in line with our previous data in LPS-activated primary microglial cells, where a synergistic inhibitory effect of ascorbic acid and aspirin on COX-2-mediated $PGE_2$ production was observed (Fiebich et al., 2003).

Cultures of the human neuroblastoma cell line SK-N-SH are an useful model to study the induction and inhibition of neuronal prostaglandin synthesis, since they express mPGES-1 and COX-2 mRNA and protein and show $PGE_2$ production upon stimulation with IL-1β (Fiebich et al., 2000b). We have previously demonstrated that IL-1β-mediated $PGE_2$ production is mainly dependent on the activity of COX-2, based on the potent inhibitory effects of highly selective COX-2 inhibitors on IL-1β-induced $PGE_2$ biosynthesis (Fiebich et al., 2000b). Moreover, the failure of the highly selective COX-1 inhibitor SC-560 to reduce COX activity (Table 1) provides further support to our previous observations, indicating the key contribution of COX-2, rather than COX-1, to the $PGE_2$ production induced by IL-1β in SK-N-SH cells.

Reduction of $PGE_2$ by ascorbic acid was not related to modification of COX-1 protein expression, which is not induced by IL-1β in SK-N-SH cells (Fig. 4A and Fiebich et al., 2000b), and which basal levels are not affected by ascorbic acid (Fig. 4A). We did not find a reduction in the amount of COX-2 protein when adding ascorbic acid to the cultures, which is in agreement with findings in a macrophage cell line, where no effect of ascorbic acid and aspirin on COX-2 expression was shown (Abate et al., 2000). Similarly, ascorbic acid alone or in combination with aspirin failed to modify

IL-1β-induced mPGES-1 expression. Although previous studies have shown that mPGES-1 is the most important isozyme involved in PGE$_2$ production under inflammatory conditions (Murakami et al., 2000; Mancini et al., 2001; Murakami et al., 2002), we can not rule out that the effects observed in the present study result from changes in the expression of other PGES isoforms (e.g., cPGES or mPGES-2) following incubation of SK-N-SH cells with IL-1β in the presence of ascorbic acid and/or aspirin.

Our data demonstrate that ascorbic acid inhibits COX-2-mediated PGE$_2$ production. One possibility is that ascorbic acid acts on the catalytic site of COX-2, which is the primary mechanism of action of aspirin-like drugs (Luong et al., 1996). Nevertheless, we excluded this possibility in the present study since neither ascorbic acid nor its oxidized metabolite DHA was able to inhibit neuronal COX enzymatic activity (Table 1). This finding is in contrast with our previous observations on the direct inhibitory effects of ascorbic acid on COX-2 activity in LPS-treated primary rat microglia (Fiebich et al., 2003). The apparent discrepancies between these studies might be due to different cell type (primary rat microglia vs. human neuroblastoma cell line). However, up to now, inhibition of the catalytic site of COX-2 has not been shown to be cell type specific for any drug. In addition, there might be differences in the expression of ascorbic acid transporters between microglial and neuronal cells, although further studies need to be performed to address this issue.

The effects of ascorbic acid on COX-2-mediated PGE$_2$ production may be mediated through its interaction with the peroxidation site of COX-2, which has also been proposed for paracetamol (Fiebich et al., 2000a; Boutaud et al., 2002). Most interestingly, the indirect inhibition of COX-2 by reducing the higher oxidative states has been shown to be cell type specific which may not only explain the cell type selective effects of paracetamol (Boutaud et al., 2002), but also the effects of the antioxidants ascorbic acid and Trolox C observed in the present study (Fig. 1). This notion is supported by our findings that ascorbic acid as well as Trolox C dose-dependently reduced IL-1β-induced neuronal production of 8-*iso*-PGF$_{2α}$ and PGE$_2$ (Figs. 1 and 6A). There is a large body of experimental evidence indicating that COX catalysis involves radical intermediates, which is further supported by the results showing that COX activity is reduced by a variety of antioxidants (Takeguchi and Sih, 1972; Vanderhoek and Lands, 1973; Hemler and Lands, 1980; Ciuffi et al., 1996; Jiang et al., 2000; Beharka et al., 2002). The present finding that the selective COX-2 inhibitor SC-58125 showed a high potency in reducing IL-1β-mediated formation of 8-*iso*-PGF$_{2α}$ (Fig. 6C), suggests that COX-2 activity is one of the major sources of ROS in neuroinflammation. Similarly, our recent results in LPS-activated microglia also showed a COX-2-dependent formation of 8-*iso*-PGF$_{2α}$ (Akundi et al., 2005). Although isoprostanes are non-enzymatic products formed by free radical attack on arachidonic acid, *in vitro* and *in vivo* evidences suggest that 8-*iso*-PGF$_{2α}$ could in part derive from COX activity (Pratico et al., 1995; Pepicelli et al., 2002). This is further supported by *in vivo* findings on the contribution of COX-2 to oxidative damage during brain ischemia (Candelario-Jalil et al., 2003a, 2003b), and traumatic brain injury (Tyurin et al., 2000), conditions in which the neuroinflammatory mechanisms play a key role in neuronal damage.

Our present observations on the effects of the antioxidants ascorbic acid and Trolox C as well as the COX inhibitors indomethacin and SC-58125 on PGE$_2$ production and free radical formation suggest that ROS are partly responsible for the regulation of COX-2-mediated PGE$_2$ production in IL-1β-stimulated neuronal cells. This might lead to a positive feedback mechanism in which COX-2 activity is induced by ROS, but at the same time COX-2 activity modulates the cellular redox status by further increasing ROS formation.

Although we investigated several possible mechanisms in an effort to explain the synergistic effects of the combination of ascorbic acid and aspirin on the reduction of PGE$_2$ in neuronal cells, the exact mechanism of such interaction is not completely clear, especially due to the fact that ascorbic acid did not synergize with other COX inhibitors used as controls (Fig. 3). This is a completely unexpected finding. The reason for the possible synergistic effects observed in this study with the combination of antioxidants (ascorbic acid or Trolox C) and aspirin (Figs. 1 and 2), but not with other NSAIDs, is probably the result of more complex redox-sensitive mechanisms involved in the particular interaction of aspirin with COX.

One important difference between aspirin and the other COX inhibitors used in our study is its unique mechanism of action, which involves the covalent modification of COX. The interactions of aspirin with COX isoforms are particularly interesting. Aspirin irreversibly inhibits COX-1 by acetylating an active site serine which resides in the arachidonic acid binding pocket (DeWitt et al., 1990). Acetylation appears to cause steric hindrance to arachidonic acid binding, thereby blocking COX-1 activity (DeWitt et al., 1990). Aspirin also causes a time-dependent acetylation of serine 516 of the human COX-2, converting this enzyme to a form which is still capable of oxygenating arachidonic acid leading to the synthesis of 15-$R$-hydroxyeicosatetraenoic acid (15-$R$-HETE) instead of $PGH_2$ (Lecomte et al., 1994). Thus, the loss of cyclooxygenase activity in aspirin-treated COX-2 is accompanied by an increase in lipoxygenase activity (Lecomte et al., 1994; Xiao et al., 1997). Unlike other NSAIDs, aspirin does not influence the peroxidase activity of COX-2 (Xiao et al., 1997; Smith et al., 2000), which converts $PGG_2$ into $PGH_2$. Therefore, after aspirin treatment, the unchecked peroxidase activity of COX-2 can continue to generate free radical species and $PGH_2$ (Smith and Marnett, 1991; Mancini et al., 1997; Xiao et al., 1997). At low doses of aspirin, which are known to more effectively inhibit COX-1 rather than COX-2 (Demasi et al., 2000), large amounts of $PGG_2$ (formed by the cyclooxygenase activity of COX-2) are then converted into $PGH_2$ by the peroxidase activity of COX-2, and further metabolized by the terminal prostaglandin synthases. At this point, antioxidants through a mechanism involving maintenance of the redox status, could indirectly reduce the peroxidase activity of COX-2, with the concomitant reduction of PGE$_2$ formation, as observed in our study (Figs. 1 and 2). This notion is further supported by the fact that the synergistic effects of ascorbic acid and aspirin are much more evident at low doses of aspirin (Fig. 1). When present at higher doses, aspirin-mediated acetylation of COX-2 results in the conversion of PGG$_2$ into 15-*R*-HETE by the lipoxygenase activity of the acetylated COX-2.

Unlike aspirin, the other NSAIDs tested in our investigation act through a mechanism involving competition with arachidonic acid for binding to COX (DeWitt, 1999; Smith et al., 2000), thus producing an inhibition of both activities of COX (cyclooxygenase and peroxidase). This probably explains why ascorbic acid does not synergize with the other NSAIDs.

Results presented in Fig. 3, which show the failure of ascorbic acid to potentiate the inhibitory effects of other NSAIDs, exclude the possibility that the effects observed with the combination of ascorbic acid and aspirin might be related to reduction in the availability of arachidonic acid or enhancement of the catabolism of PGE$_2$ mediated by ascorbic acid.

It is noteworthy that ascorbic acid reduced IL-1β-mediated formation of ROS and PGE$_2$ in neuronal cells since this cytokine is one of the key player of brain as well as peripheral inflammation (for reviews see Rothwell and Luheshi, 2000; Basu et al., 2004; Braddock and Quinn, 2004). Although DHA has no antioxidant capacity by itself, it was able to reduce 8-isoprostane formation at the highest dose (Fig. 6A). A plausible explanation for this finding might be that DHA is recycled intracellularly to ascorbic acid and that it is the latter which might produce the observed inhibitory effect. DHA is physiologically formed either directly or via the disproportionation of the ascorbyl radicals and this oxidised form of vitamin C is directly converted to ascorbic acid either by direct chemical reaction with reduced glutathione (GSH), which is believed to be the major pathway for ascorbic acid recycling (Winkler et al., 1994), or by NAD(P)H-dependent enzymes such as thioredoxin reductase and GSH-dependent DHA reductase (Wells and Xu, 1994; May et al., 1997; May, 2002). Although DHA can be recycled into ascorbic acid, it is rapidly degraded/metabolized; thus resulting in lesser effective concentration of ascorbic acid in the cell than the tested dose.

The inhibitory effects of ascorbic acid alone and the possible synergistic effects when combined with aspirin might be of clinical relevance for diseases of the CNS when neuronal COX-2 inhibition is the therapeutic target. Oral dosing with ascorbic acid results in plasma concentrations that reach a plateau of 80 μM, and intracellular concentrations reach a plateau in the range of 1.4 to 3.4 mM with a dose of 200 mg per day (Levine et al., 1996). Addition of ascorbic acid, which is extracted from the blood stream and concentrated in the CNS to high levels (Rice and Russo-Menna, 1998), may augment the inhibition of neuronal COX-2 by aspirin especially in elderly people, in which cerebral levels of ascorbic acid may decline (Svensson et al., 1993; Miele and Fillenz, 1996; Rice and Russo-Menna, 1998).

Inhibition of the neuronal COX-2 enzyme activity may be a therapeutic target in AD and cerebral ischemia (Iadecola and Alexander, 2001; Pasinetti, 2001; Hull et al., 2002; Candelario-Jalil et al., 2004). Prospective epidemiological data repeatedly confirmed a strong protective effect of NSAIDs (in t'Velt et al., 2001) and a small protective effect of ascorbic acid (Engelhart et al., 2002) against the development of AD. Clinical studies with NSAIDs for inhibiting COX in the CNS have mainly been hampered by peripheral toxicity (Rogers et al., 1993). After 1995, a new class of selective COX-2 inhibitors (e.g., rofecoxib, celecoxib) was widely marketed in an effort to reduce side effects associated to the use of non-selective COX inhibitors. However, recent evidences indicate that selective COX-2 inhibitors are unsafe for clinical use, with patients having an increased risk of cardiovascular events (Juni et al., 2004). Preferential inhibition of neuronal COX-2 by addition of ascorbic acid to a lower dose of aspirin may be helpful to overcome peripheral side effects.

Furthermore, recent studies show that ascorbic acid attenuates aspirin-induced gastric damage (Brzozowski et al., 2001; Pohle et al., 2001; Becker et al., 2003; Dammann et al., 2004; Konturek et al., 2004).

In summary, the key finding of this investigation is our observation that ascorbic acid substantially enhances COX-2 inhibition by aspirin. At low aspirin doses that only partially reduced $PGE_2$ formation, simultaneous addition of ascorbic acid virtually abolished IL-1β-induced $PGE_2$ production. This finding may provide an important *in vitro* evidence for clinical observations showing that concurrent administration of ascorbic acid allows for dose reduction of aspirin without reducing its therapeutic efficacy. This study provides the first evidence for a possible synergism of aspirin and ascorbic acid in neuronal cells, which may have important implications for the treatment of neuroinflammatory conditions. According to the results presented here, co-administration of ascorbic acid constitutes a novel approach to render COX-2 more sensitive to inhibition by aspirin. Thus, anti-inflammatory therapy might be successful with lower doses of aspirin when combined with ascorbic acid, thereby avoiding the side effects of the usually required high dose aspirin treatment, and adding the antioxidant benefits of ascorbic acid.

**Acknowledgements:** We thank Dr. Uwe Gessner for helpful discussion**.** The skillful technical assistance of Ulrike Götzinger-Berger, Franziska Klott, Brigitte Günter, and Sandra Hess is greatly acknowledged. This work was supported in part by grant Fi 683/1-1 from the Deutsche Forschungsgemeinschaft. ECJ was supported by a research fellowship from the Alexander von Humboldt Foundation (Bonn, Germany).

**Table 1**

Effects of ascorbic acid alone or in combination with aspirin on COX enzymatic activity in SK-N-SH cells.

| Treatment | PGE$_2$ (% of IL-1β + Arachidonate Control) |
|---|---|
| ***Effect of ASA and the combination of ASA and Ascorbic acid on COX enzymatic activity*** | |
| Control | 0.4 ± 0.2 |
| Control + Arachidonate 15 µM | 4.1 ± 1.8 |
| IL-1β (10 U/ml) without Arachidonate | 7.5 ± 1.8 |
| IL-1β (10 U/ml) + Arachidonate 15 µM | 100 ± 9.6 |
| IL-1β + Arachidonate + ASA 1 µM | 91.0 ± 9.2 |
| IL-1β + Arachidonate + ASA 1 µM + Ascorbic acid 10 µM | 81.4 ± 13.4 |
| IL-1β + Arachidonate + ASA 1 µM + Ascorbic acid 100 µM | 98.3 ± 3.8 |
| IL-1β + Arachidonate + ASA 10 µM | 77.3 ± 7.2* |
| IL-1β + Arachidonate + ASA 10 µM + Ascorbic acid 10 µM | 80.1 ± 5.0*, § |
| IL-1β + Arachidonate + ASA 10 µM + Ascorbic acid 100 µM | 84.0 ± 1.8*, § |
| IL-1β + Arachidonate + ASA 100 µM | 52.1 ± 2.8** |
| IL-1β + Arachidonate + ASA 100 µM + Ascorbic acid 10 µM | 56.3 ± 6.7**, § |
| IL-1β + Arachidonate + ASA 100 µM + Ascorbic acid 100 µM | 58.8 ± 5.9**, § |
| ***Effect of Ascorbic acid on COX enzymatic activity*** | |
| IL-1β + Arachidonate + Ascorbic acid 10 µM | 86.4 ± 13.4 |
| IL-1β + Arachidonate + Ascorbic acid 100 µM | 89.1 ± 18.3 |
| ***Effect of COX inhibitors (control experiment)*** | |
| IL-1β + Arachidonate + SC-58125 0.1 µM | 46.4 ± 4** |
| IL-1β + Arachidonate + SC-58125 1 µM | 33.7 ± 3.9** |
| IL-1β + Arachidonate + Indomethacin 0.1 µM | 10.6 ± 5.9*** |
| IL-1β + Arachidonate + Indomethacin 1 µM | 1.5 ± 1.3*** |
| IL-1β + Arachidonate + SC-560 10 nM | 99.9 ± 7.6 |
| IL-1β + Arachidonate + SC-560 100 nM | 86.1 ± 15 |

Data are expressed as mean ± S.D. Cells were either left untreated or were stimulated with IL-1β (10 U/ml) for 24 h. After removal of medium, cells were treated with different concentrations of the compound for 30 min in absence or presence of arachidonic acid. PGE$_2$ in the supernatants was determined by an enzyme immunoassay as described in Materials and Methods. *p<0.05, **p<0.01 and ***p<0.001 with respect to IL-1β + Arachidonate control. §p>0.05, no statistically significant difference with respect to the corresponding dose of aspirin alone.

**Captions for Figures**

**Fig. 1**. Inhibitory effects of ascorbic acid (ASC) and aspirin (ASA) (**Panel A**) and Trolox C and Indomethacin (**Panel B**) on prostaglandin $E_2$ synthesis in IL-1β-stimulated human SK-N-SH cells. Cells were pre-incubated with the inhibitors at the indicated concentrations for 30 min and subsequently treated with IL-1β (10 U/ml) for 24 h. $PGE_2$ in the supernatants was measured by an enzyme immunoassay as described in Materials and Methods. Data are expressed as mean ± S.D. In Fig. 1A, the label of the X axis refers to the concentration of either aspirin or ascorbic acid when each drug is used alone, and at the same time, the X axis refers to aspirin concentration when used in combination with the given dose of ascorbic acid, as explained in the inset legend. *$p<0.05$ and **$p<0.01$ with respect to IL-1β-treated control.

**Fig 2**. The potent antioxidant Trolox C, a synthetic analogue of α-tocopherol, enhances the inhibitory effect of aspirin on $PGE_2$ production following stimulation with IL-1β in neuronal cells. Cells were pre-incubated with the inhibitors at the indicated concentrations for 30 min and subsequently treated with IL-1β (10 U/ml) for 24 h. $PGE_2$ in the supernatants was measured by an enzyme immunoassay as described in Materials and Methods. *$p<0.05$ with respect to IL-1β-treated control. **$p<0.01$ with respect to IL-1β control and aspirin alone. § $p<0.05$ with respect to Trolox C (10 μM) and IL-1β-treated control. # $p<0.01$ with respect to Trolox C (10 μM) and IL-1β-treated control. Statistical analysis was performed using one-way ANOVA followed by Student-Newman-Keuls post-hoc test. Data are expressed as mean ± S.D.

**Fig. 3.** Lack of additive effect of ascorbic acid with the highly COX-2 inhibitor SC-58125 (**A**). Similarly, ascorbic acid (ASC) did not synergize with the non-selective COX inhibitor indomethacin (**B**). Cells were pre-incubated with the inhibitors at the indicated concentrations for 30 min and subsequently treated with IL-1β (10 U/ml) for 24 h. $PGE_2$ in the supernatants was measured by an enzyme immunoassay. *p<0.05 and **p<0.01 with respect to IL-1β-treated control. Statistical analysis was performed using one-way ANOVA followed by Student-Newman-Keuls post-hoc test. Data are expressed as mean ± S.D.

**Fig. 4.** IL-1β-induced synthesis of COX-2 in SK-N-SH cells is not affected by ascorbic acid and aspirin alone or in combination (**Panel A**). Cell lysates from untreated control cell cultures or cell cultures treated for 24 h with the indicated amounts of IL-1β (50 U/ml) alone or in combination with ascorbic acid and aspirin were separated by SDS-PAGE on a 7.5% acrylamide gel, blotted and incubated with antibodies against COX-1, COX-2 and actin. Densitometric analysis of the blots is presented in **Panel B**.

**Fig. 5**. Lack of effect of the combination of ascorbic acid and aspirin (**A**), different doses of ascorbic acid (**B**) and dehydroascorbic acid (DHA, **C**) on microsomal prostaglandin E synthase-1 (mPGES-1) expression. SK-N-SH cells were treated with IL-1β (10 U/ml) for 4 h in the absence or presence of different concentrations of ascorbic acid, aspirin or DHA. Total RNA was extracted and subjected to RT-PCR.

**Fig. 6.** Effect of ascorbic acid (ASC), dehydroascorbic acid (DHA), Trolox C (**Panel A**), aspirin and the combination of aspirin with ascorbic acid (**Panel B**), as well as the COX inhibitors indomethacin and SC-58125 (**Panel C**) on 8-isoprostaglandin $F_{2\alpha}$ (8-*iso*-$PGF_{2\alpha}$) production in response to 10 U/ml IL-1β. 8-*iso*-$PGF_{2\alpha}$ levels were quantitated in the culture medium 24 h following stimulation of SK-N-SH cells with IL-1β alone, or in combination with the substances at the given concentrations. * $p<0.05$ and ** $p<0.01$ with respect to IL-1β alone. $^{\S}$ $p<0.05$ with respect to aspirin alone, but not different from ascorbic acid alone. $^{\#}$ $p<0.05$ with respect to ascorbic acid alone, and the corresponding dose of aspirin without ascorbic acid. Histogram represents mean ± S.D. of 3 independent experiments. Statistical analysis was performed using one-way ANOVA followed by Student-Newman-Keuls post-hoc test.

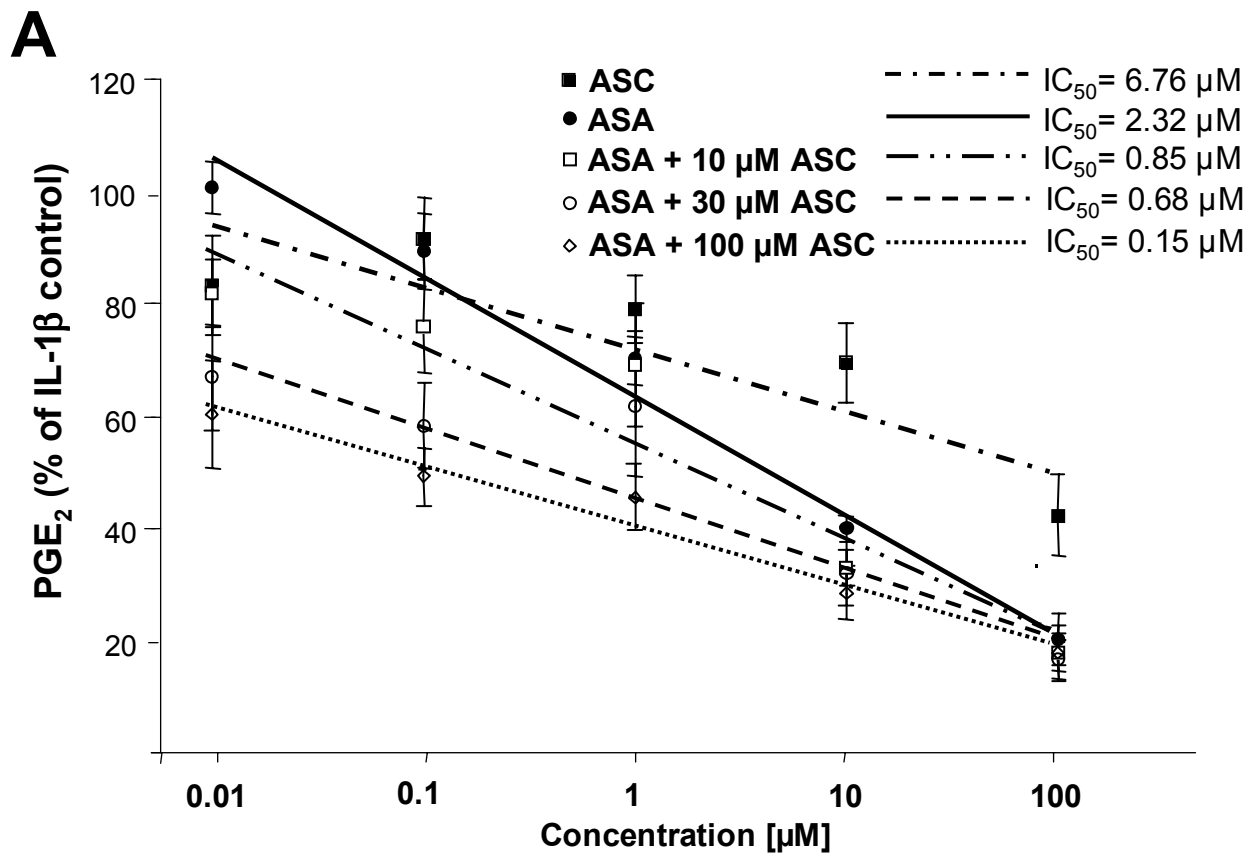

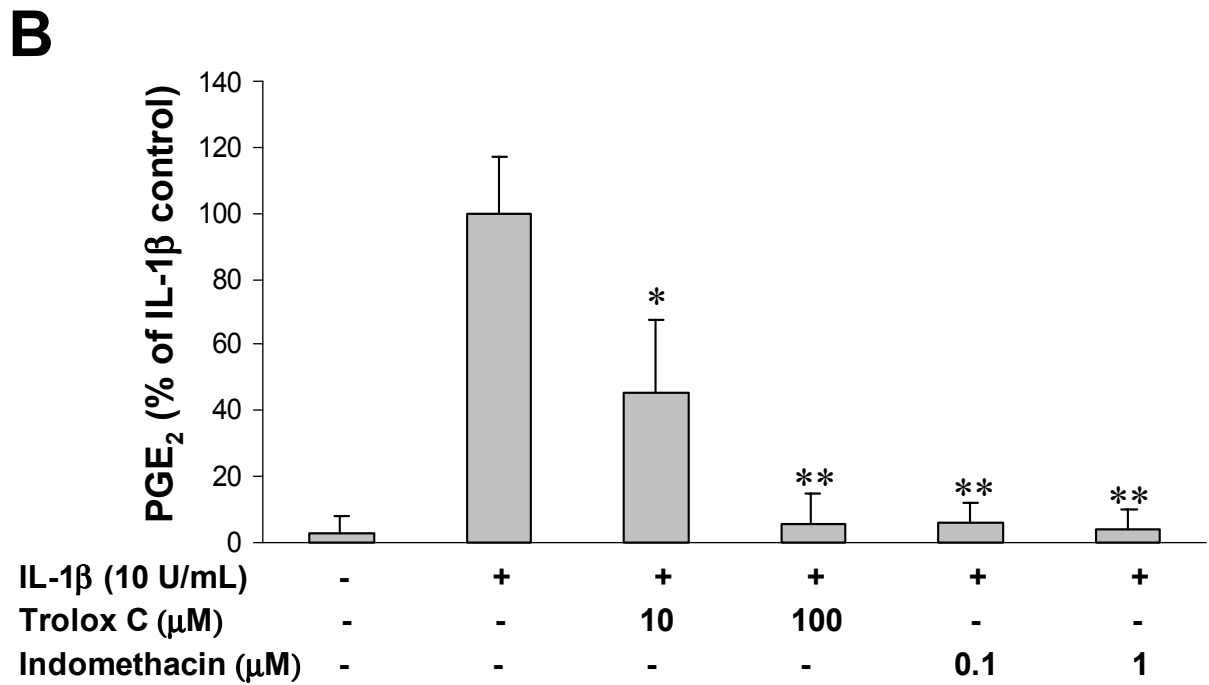

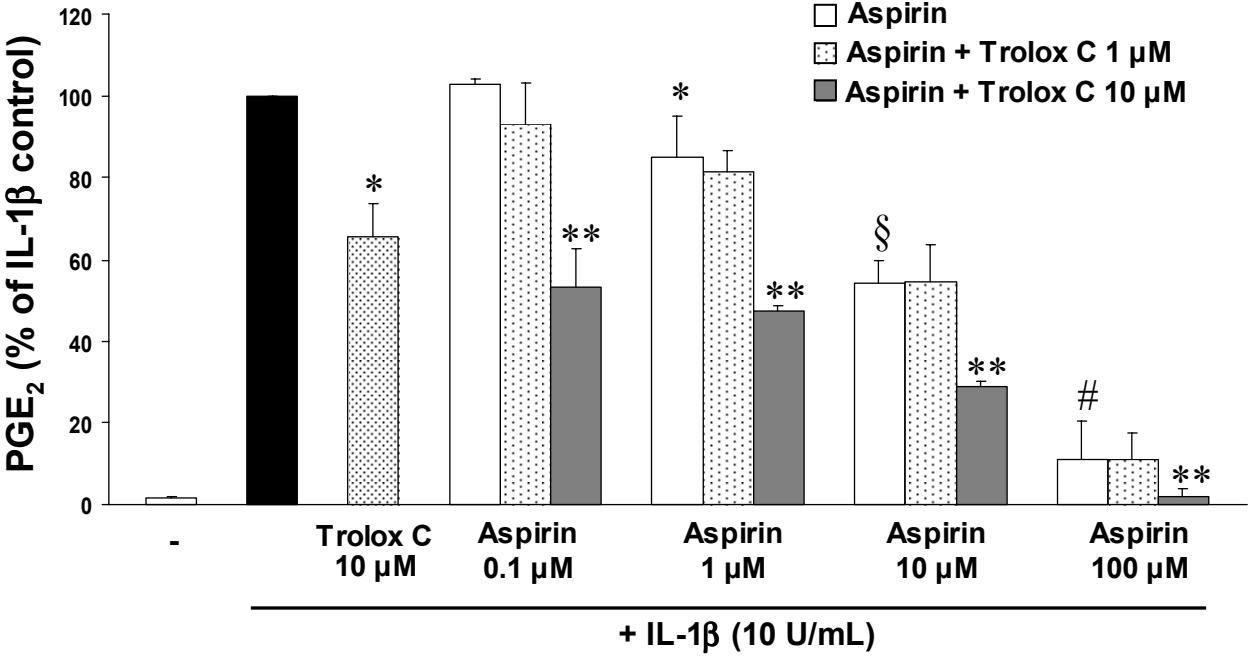

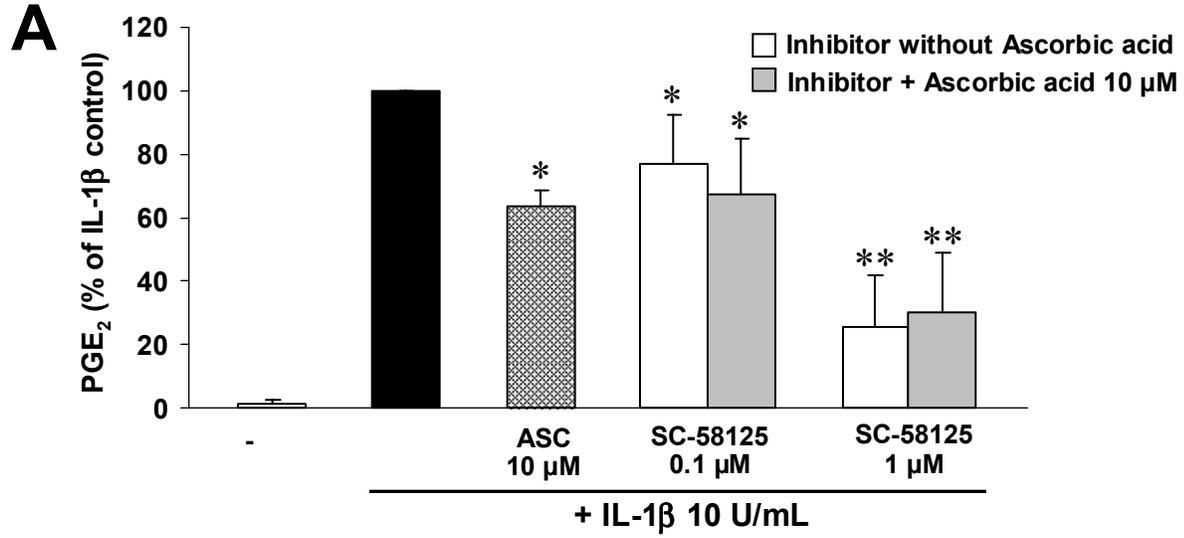
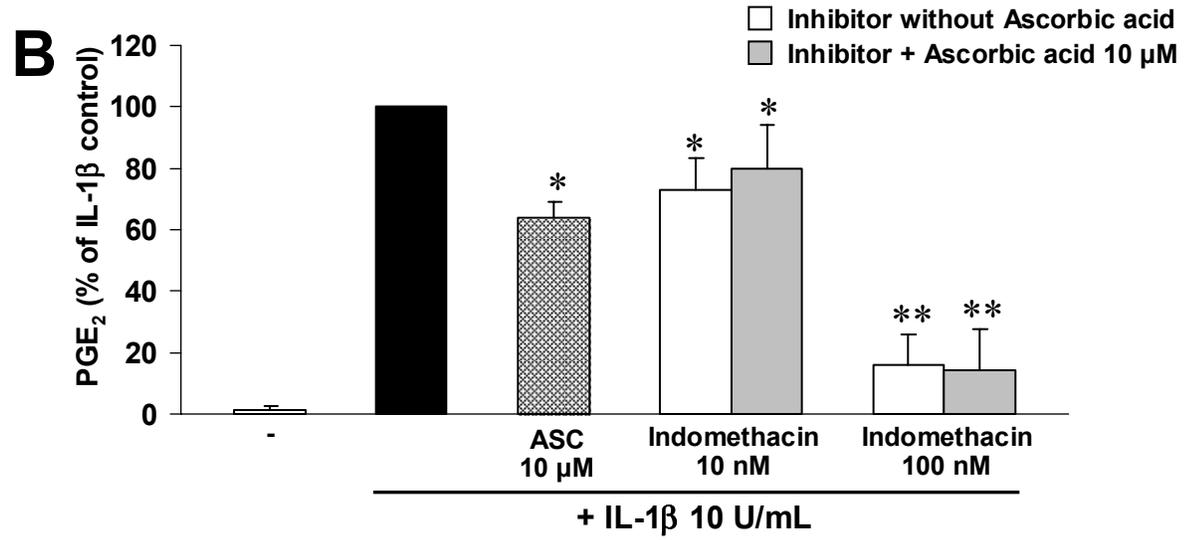

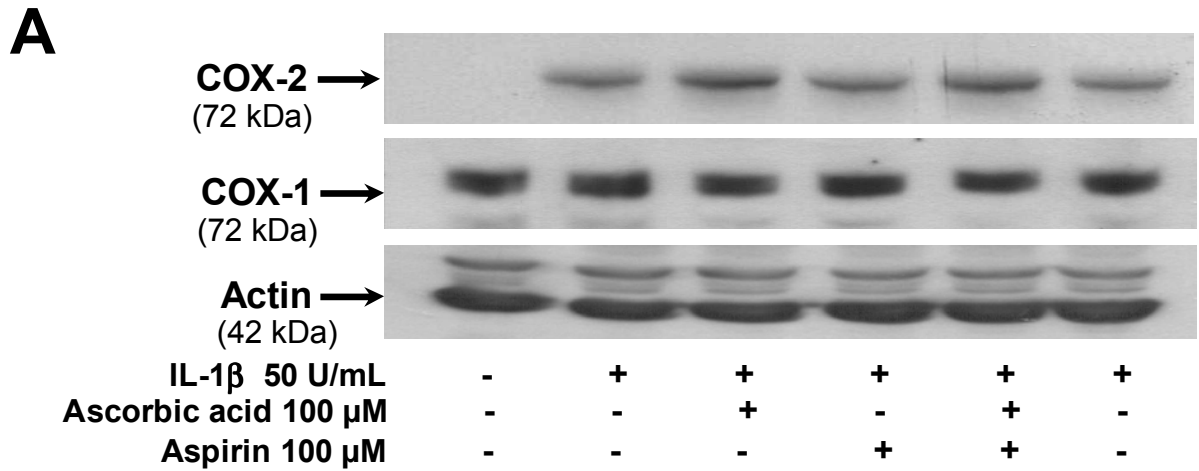

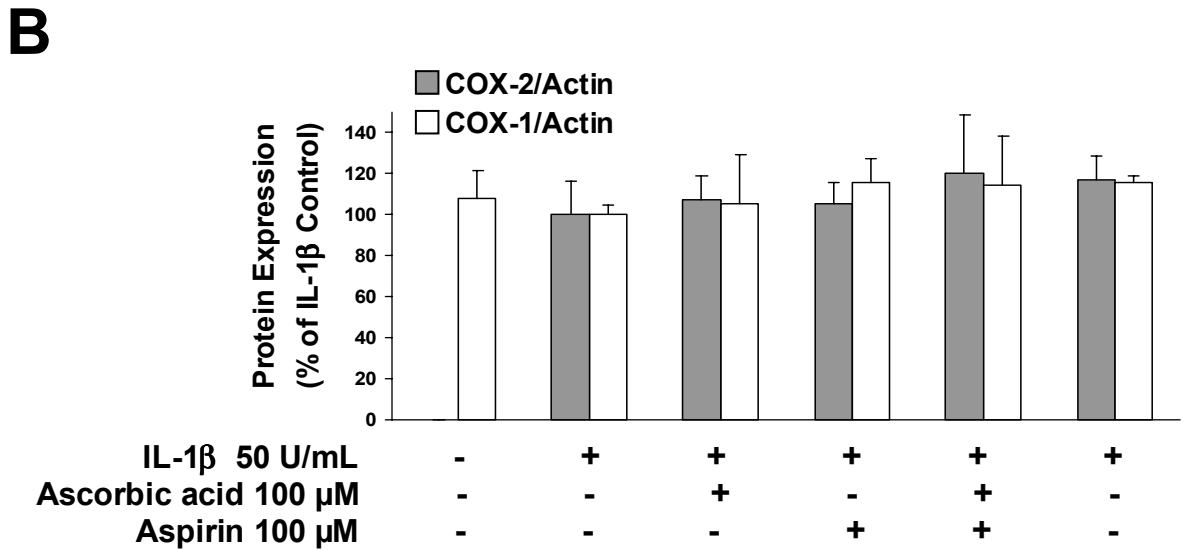

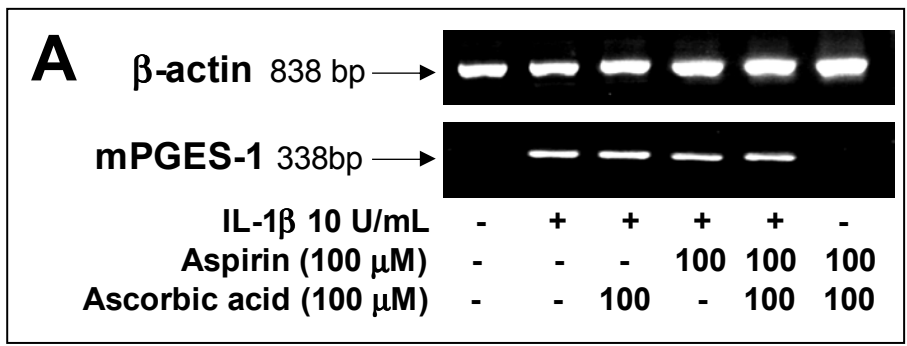
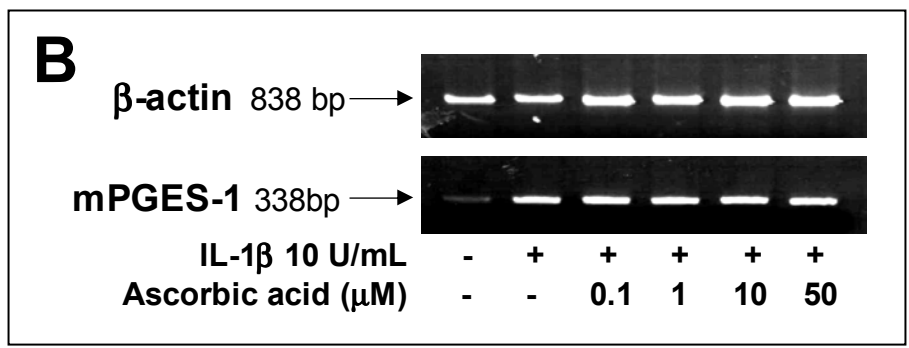
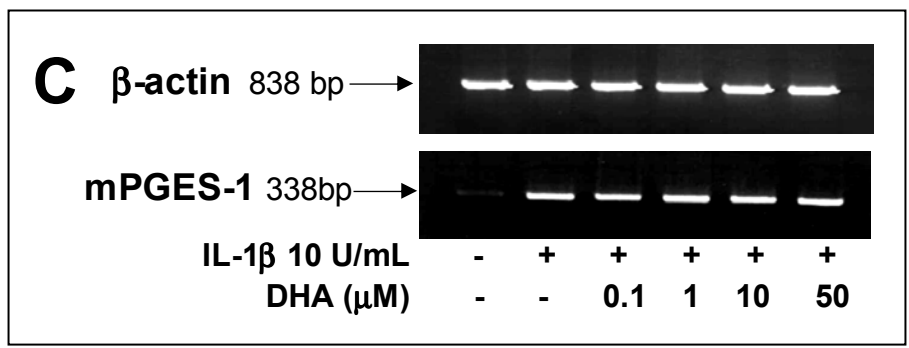

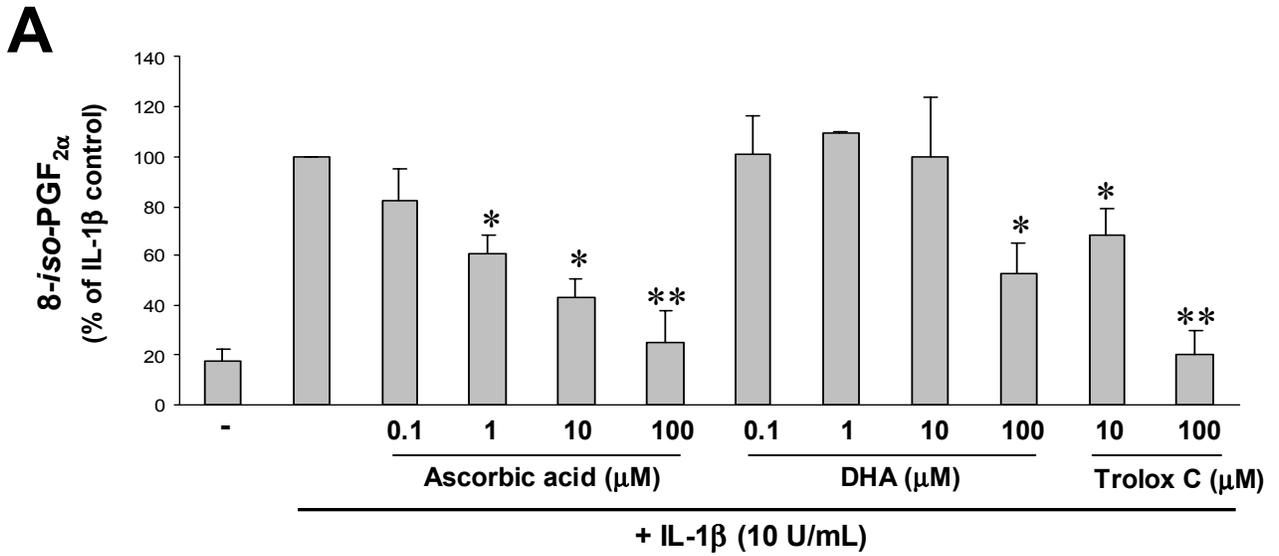
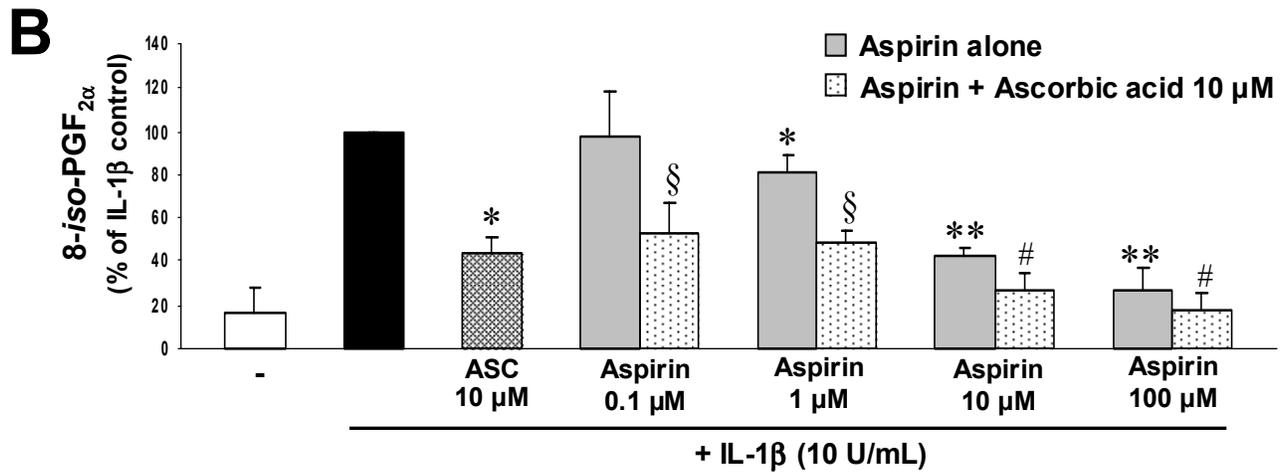
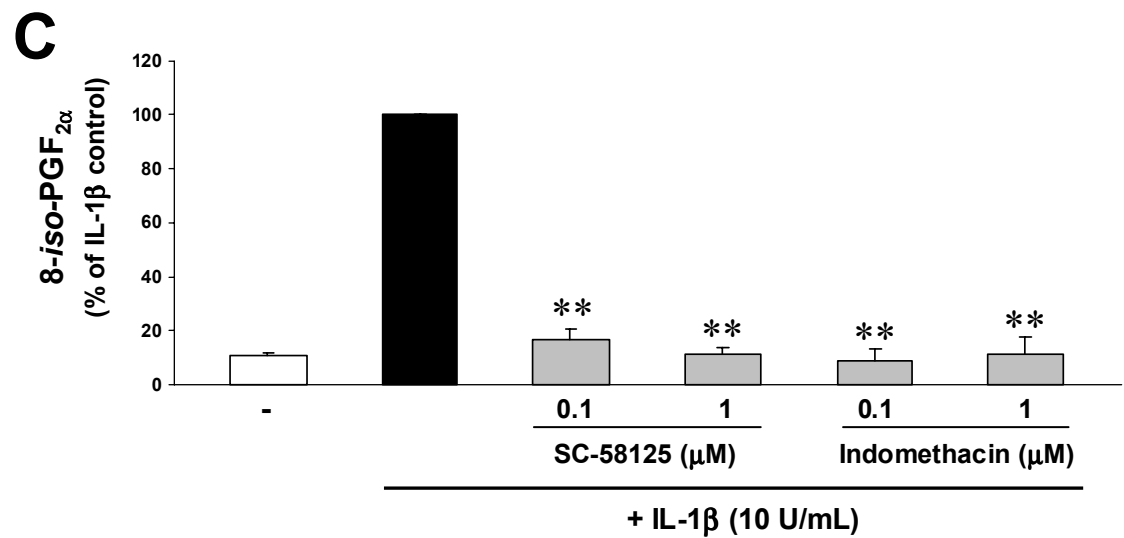